\documentstyle[prd,epsfig,aps,preprint]{revtex}
\begin{document}

\draft
\title{On the Width of the Last Scattering Surface}

   \author{N. PIRES\footnote{npires@dfte.ufrn.br}, M. A. S. NOBRE\footnote{assunta@fisica.ufpb.br},
   J. A. S. LIMA\footnote{limajas@astro.iag.usp.br}
          }

  \smallskip
\address{~\\Departamento de F\'{\i}sica, \\Universidade Federal
             do Rio Grande do Norte, C.P. 1641, 59072-970, Natal,
             RN, Brasil
                  }

\date{\today}
\maketitle


\begin{abstract}We discuss the physical effects of some accelerated world models
on the width of the last scattering surface (LSS) of the cosmic
microwave background radiation (CMBR). The models considered in
our analysis are X-matter (XCDM) and a Chaplygin type gas.  The
redshift of the LSS does not depend on the kind of dark energy (if
XCDM of Chaplygin). Further, for a Chaplygin gas, the width of the
LSS is also only weakly dependent on the kind of scenario (if we
have dark energy plus cold dark matter or the unified picture).
\end{abstract}


\section{Introduction}

It is widely known that the last scattering surface (LSS) has a
finite thickness $\Delta z$ (in redshift space) because the
hydrogen recombination process takes a finite time. This fact has
important consequences for the physics relating theoretical
predictions and observations of the the angular pattern of CMBR
anisotropies. In particular, it implies that anisotropies at
length scales smaller than $\Delta z $ must be naturally
suppressed.

On the other hand, recent observations from type Ia
Supernovae\cite{1} combined with CMBR experiments\cite{2} strongly
suggest an accelerating and nearly flat Universe ($q_o < 0,\
\Omega_{Total}\approx 1$). In the framework of general relativity
both facts are readily accommodated by assuming the existence of
an extra dark energy (DE) component with negative pressure (in
addition to cold dark matter (CDM)). Besides the cosmological
constant\cite{Peebles1} ($\Lambda$) and a scalar field\cite{PAD}
($\Phi$), there are at least 3 distinct dark energy candidates
proposed in the literature, namely: vacuum decay
$\Lambda(t)-models $\cite{Lima}, X-matter\cite{X}, and a Chaplygin
gas\cite{CG}.

In this paper we discuss the width of the LSS for accelerating
world models driven by the last two above quoted DE candidates
(X-matter and Chaplygin gas). Since the dark energy component
present in these models is separately conserved, the consequences
to LSS are completely different to what happens with accelerating
models endowed with any kind of adiabatic photon
creation\cite{Lima,mc}. With basis on the WMAP
observations\cite{W}, in the present calculations we consider
$H_0= 71$ km s$^{-1}$Mpc$^{-1}$, $\Omega_b=0.04$ for baryonic
matter, and $\Omega_{DM}=0.27$ for cold dark matter.

\section{Last Scattering Surface and the Recombination Process}

The probability that a CMBR photon undergone its last scattering
between $z$ and $z+ dz$ is $P = 1 - e^{-\tau (z)}$, where $\tau =
\int\limits_0^z n_e\; \sigma_T\; c\; dt$ is the optical depth to
redshift $z$, $\sigma_T$ is the Thomson cross section and $n_e$ is
the density of electrons. The quantity ${dP(z)}/{dz}$ determines
the visibility function, that is, the probability distribution for
the redshifts where the CMBR photons had their last scattering. By
defining the effective profile of the LSS:
$V(z)=dP(z)/dz=e^{-\tau(z)}d\tau/dz$, and fitting the visibility
curve with a Gaussian form, the standard deviation yields a
reasonable estimate of the LSS width whereas its peak stands for
the beginning of the recombination epoch\cite{4}.

At least 3 physical process are acting during and after
recombination on the baryonic matter: photoionization, cooling
from recombination, and Compton cooling-heating. Hence, neglecting
the helium and treating the hydrogen atom as a two-level system,
the fraction $x_e$ of ionized matter obeys\cite{5}
\begin{equation}
{\frac{dx_e}{dt}=\frac{\Lambda_{2s,1s}}{(\Lambda_{2s,1s}+\beta_e)}\left[\beta_e
e^{-\frac{(B_1-B_2)}{k_\beta T_\gamma}}(1-x_e)-\frac{a_r \rho
x_e^2}{m_p}\right]}\, , \label{xe}
\end{equation}
where $B_1=13.6\ {\mathrm eV}$ is ground state energy, $B_2=3.4\
{\mathrm eV}$ is the first excited state energy, $\beta_e = (2\pi
m_ek_{{}_B}T_\gamma)^{3/2}h^{-3} e^{-(B_2/k_{{}_B}T_\gamma)}a_r\;$
is the photoionization rate, $a_r=2.84\times 10^{-11}\;
T_m^{-1/2}$cm$^3$ sec$^{-1}$ is the recombination coefficient, and
$\Lambda_{2s,1s}=8.272 $ sec$^{-1}$ is the two-photon emission
rate.
After decoupling, the matter temperature ($T_m$) of the neutral
atoms fall faster than the radiation temperature ($T_\gamma$). The
matter temperature decreasing is governed by the equation
\begin{equation}
\frac{dT_{m}}{dt} = T_{m}\left[ -2\frac{\dot R}R - \frac{\dot
x_e}{3\left( 1+x_e\right) } \right]  - \frac{8\sigma_T b}{3m_e
c}\frac{T_\gamma^4 x_e} {\left( 1+x_e\right) }\left(T_{m}-T_\gamma
\right) \, , \label{Tm}
\end{equation}
where $\dot R/R$ is the  Hubble parameter, $\dot x_e = dx_e/dt$,
and $8\sigma_T b/3 m_e c = 8.02\times10^{-9}$ sec$^{-1}$K$^{-4}$.

\section{Visibility Function: Main Results}

Let us now discuss the visibility function for the accelerating
models (X-matter and Chaplygin gas) quoted in the introduction.

{\bf (i) X-Matter Models:}

In cosmological scenarios driven by X-matter plus cold dark matter
(sometimes called XCDM parametrization) both fluid components are
separately conserved. The equation of state of the dark energy
component is $p_x = w (z) \rho_x$. Unlike to what happens with
scalar field motivated models where $w(z)$ is derived from the
field description, the expression of $w(z)$ for XCDM scenarios
must be assumed a priori. Models with constant $w$ are the
simplest ones and their free parameters can easily be constrained
from the main cosmological tests. In what follows we focus our
attention to this class of models assuming a flat geometry. The
differential time-redshift relation is
\begin{equation}
dt=\frac{1}{H_0}\frac{dx}{ x \left[ \Omega_{M} x^{-3} + (1 -
\Omega_{M}) x^{-3(1+\omega)}\right]^{1/2} } \, , \label{quinta}
\end{equation}
where $\Omega_{M} = 1- \Omega_{X}$ is the density parameter of the
dark matter. Taking the limiting case $\omega=-1$, the
$\Lambda$CDM results are recovered. The basic results are
presented in Table 1. The left panel of Figure 1 shows the
corresponding visibility function.

{\bf (ii) Chaplygin Gas:}

This class of accelerating models refers to an exotic fluid whose
equation of state is given by  $p_{C} = -A/\rho_{C}^{\alpha}$,
where $A$ and $\alpha$ are positive parameters. Actually, the
above equation for $\alpha \neq 1$ generalizes the original
Chaplygin equation of state whereas for $\alpha = 0$, the model
behaves like scenarios with cold dark matter plus a cosmological
constant ($\Lambda$CDM). The dynamics of such a fluid is similar
to non-relativistic matter (dark matter) at high redshift and as a
negative-pressure DE component at late times. Two different
pictures are usually considered in the literature: the first is a
flat scenario driven by a non-relativistic matter plus the
Chaplygin gas as a dark energy (GCgCDM), whereas in the second
one, the Chaplygin type gas together with the observed baryonic
content are responsible by the dynamics of the present-day
universe (unifying dark matter with dark energy (UDME or
Quartessence). The differential age-redshift relation as a
function of the observable now reads
\begin{equation}
dt=\frac{1}{H_0}\left\{
\frac{x}{\Omega_{j}+(1-\Omega_{j})x^3[A_s+(1-A_s)
x^{-3(\alpha+1)}]^{\frac{1}{1+\alpha} } }\right\}^{1/2}dx \, ,
\label{Chaply}
 \end{equation}
where  $A_s=A/ \rho_{c_0}^{1+\alpha}$ and $H_{0}$ is the Hubble
constant. $\Omega_{j}$ stands for  baryonic + dark matter density
parameter in GCgCDM models but only to the baryonic matter density
parameter in the UDME (Quartessence) scenarios. The results are
presented in Table 2 and the right panel of Figure 1 show the
corresponding visibility function to both cases.

In summ, the thickness of the LSS has been discussed using two
different accelerating world models. Tables  1 and 2 show the main
conclusions of this work. As we have seen, the X-matter models
present the same behaviour of constant $\Lambda$ models,
regardless of the value of $\omega$. Probably, more important, the
recombination epoch is just the same for all models driven by
X-matter and the Chaplygin gas (it is located at redshift
$z_{rec}=1.127$). Further, the width of the LSS is only weakly
dependent on the kind of dark energy models considered here. As
one can see from Table 2, the UDME models (in which the C-gas
plays the role of both dark matter and dark energy) has a little
influence on the width of the LSS. This is in line with the
visibility function presented in figure (b).

Finally, for the sake of comparison, we have also computed
$z_{rec}$ and the width of the LSS for models with decaying vacuum
energy density\cite{Lima} and adiabatic gravitational creation of
matter and radiation\cite{mc}. Due to the adiabatic creation of
photons, the results concerning the width of the LSS and $z_{rec}$
are strongly modified. This means that the physics of the LSS may
constrain with great accuracy any model endowed with photon
creation because the temperature law of the CMBR is modified. This
problem will be discussed in a forthcoming communication.

{\bf Acknowledgements:}The authors are grateful to J. C. Neves de
Araujo by the numerical code and many helpful discussions.

\begin{table}[h]
\centerline{Redshift and width of the LSS for X-matter models}
 {\begin{tabular}{@{}cccccccc@{}}
 MODEL   &$\Omega_{M}$&$\Omega_x$ &$\lambda$ &$\omega$         &$\Delta z$&$z_{rec}$ & $\Delta$l (Mpc)\\
 X-matter&  0.27       &0.69       & --        &\hphantom{0}0.0  &68.7      &1127.4    &13.8     \\
         &  0.27       &0.69       & --        &           -0.5  &68.7      &1127.4    &13.8      \\
         &  0.27       &0.69       & --        &           -1.0  &68.7      &1127.4    &13.8      \\
$\lambda$-model&0.27 & -- & 0.69      &   -- &68.7 &1127.4 &13.8
\\
 \end{tabular}}
\end{table}

\begin{table}[h]
\centerline{Redshift and width of the LSS for a Chaplygin gas.}
 {\begin{tabular}{@{}cccccccc@{}}
 MODEL  &$\Omega_j$&$\Omega_C$&$\alpha$&$A_s$&$\Delta z$&$z_{rec}$&$\Delta$l (Mpc) \\
  GCgCDM&0.31   & 0.69 & 1.0  &0.4  & 108.8             &1127.4 & 13.2       \\
        &0.31   & 0.69 & 1.0  &0.5  & \hphantom{0}68.7  &1127.4 & \hphantom{0}8.6  \\
        &0.31   & 0.69 & 1.0  &0.7  & 108.8             &1127.4 & 14.6      \\
        &0.31   & 0.69 & 1.0  &0.9  & 108.8             &1127.4 & 16.7      \\
        &0.31   & 0.69 &  -   &1.0  & \hphantom{0}68.7  &1127.4 & 13.8      \\
        &0.31   & 0.69 & 0.8  &0.5  & 108.8             &1127.4 & 13.8      \\
        &0.31   & 0.69 & 0.5  &0.5  & 108.3             &1127.4 &14.0       \\
        &0.31   & 0.69 & 0.2  &0.5  & 108.3             &1127.4 &14.5       \\
              &&&&&&&\\
 UDME   &0.04   & 0.96 & 1.0  & 0.4 & 108.8             &1127.4 &13.7       \\
        &0.04   & 0.96 & 1.0  & 0.5 & \hphantom{0}68.7  &1127.4 &10.7       \\
        &0.04   & 0.96 & 1.0  & 0.7 & 108.8             &1127.4 &16.1       \\
        &0.04   & 0.96 & 1.0  & 0.9 & 108.8             &1127.4 &20.7       \\
        &0.04   & 0.96 &  -   & 1.0 & \hphantom{0}68.7  &1127.4 &38.3       \\
        &0.04   & 0.96 & 0.8  & 0.5 & 108.8                  &1127.4 &14.6       \\
        &0.04   & 0.96 & 0.5  & 0.5 & 108.8             &1127.4 &15.1       \\
        &0.04   & 0.96 & 0.2  & 0.5 & 108.8             &1127.4 &16.0       \\

 \end{tabular}}
\end{table}

\begin{figure}[th]
\centerline{\psfig{file=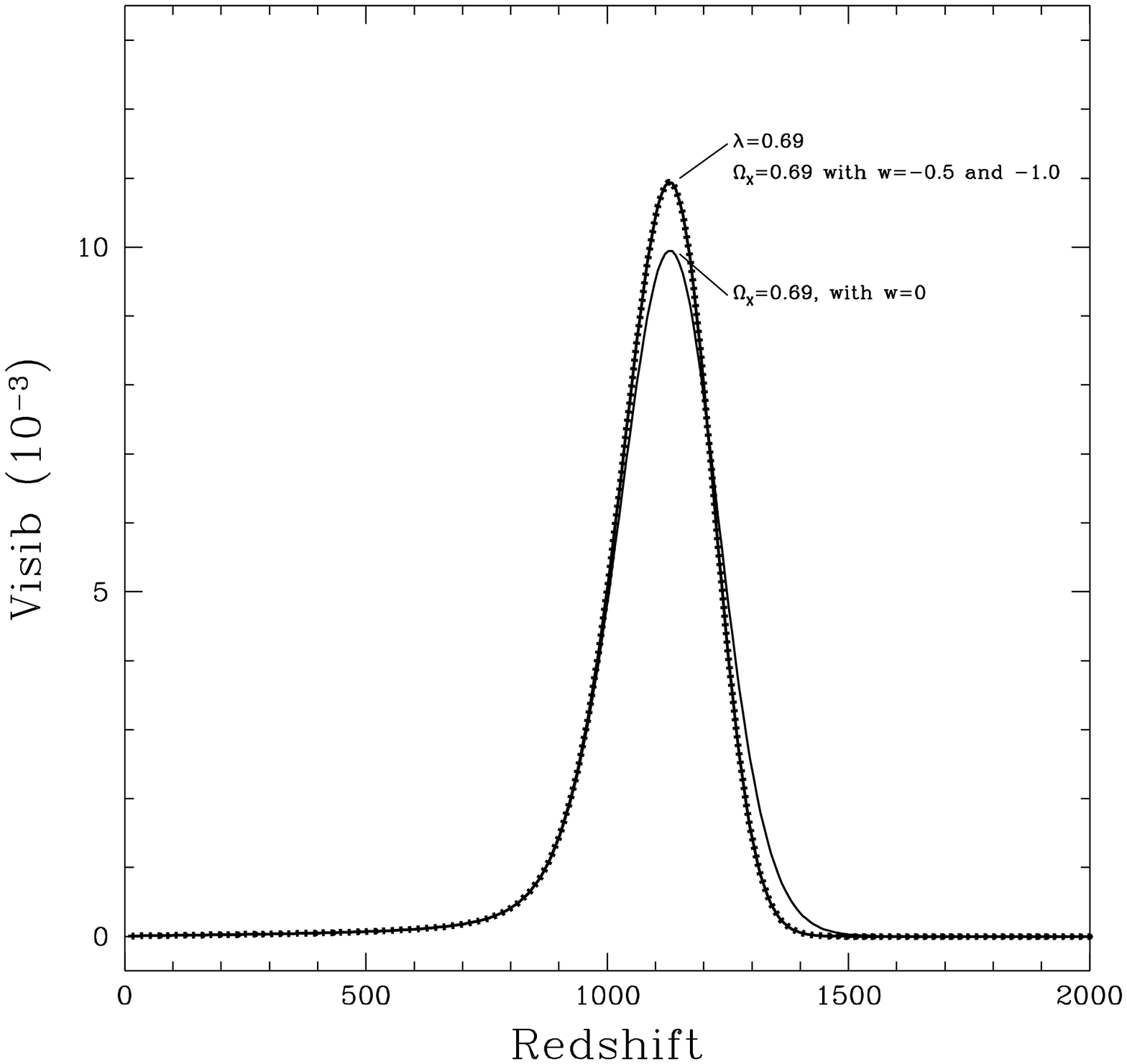,width=6.9cm}\hspace{-.7truecm}\psfig{file=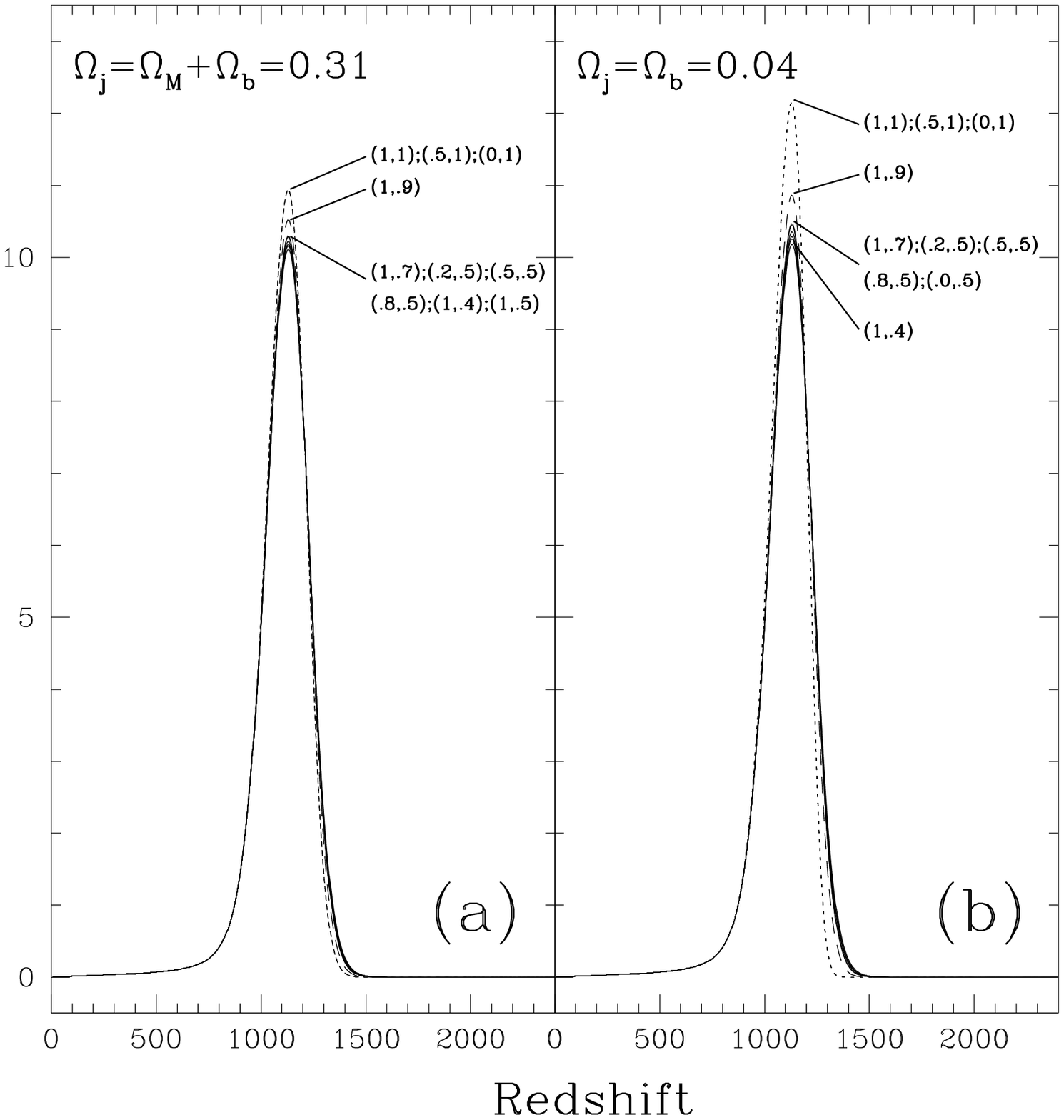,width=6.9cm}}
\vspace*{8pt} \caption{Visibility function for models with
X-matter and a Chaplygin gas. In the first panel (left), the
dotted line is the prediction for $\Lambda$CDM models whereas the
others lines are for XCDM. The two panels on the right show the
results for a Chaplygin type gas: (a) models with CDM plus a
Chaplygin gas (GCgCDM), and (b) Chaplygin gas in the unified
(UDME) scenario. Note that the visibility function is not very
affected by these models.}
\end{figure}

\end{document}